\documentclass{elsart}

\begin{document}
\setlength{\parindent}{0mm}
\setlength{\parskip}{3mm}
\setlength{\baselineskip}{0.55cm}
\def\MS{\hbox{$\overline{\rm MS}$}}
\def\PR{{\it Phys.~Rev.~}}
\def\PRL{{\it Phys.~Rev.~Lett.~}}
\def\NP{{\it Nucl.~Phys.~}}
\def\NPBPS{{\it Nucl.~Phys.~B (Proc.~Suppl.)~}}
\def\PL{{\it Phys.~Lett.~}}
\def\PRep{{\it Phys.~Rep.~}}
\def\AP{{\it Ann.~Phys.~}}
\def\CMP{{\it Comm.~Math.~Phys.~}}
\def\JMP{{\it Jour.~Math.~Phys.~}}
\def\NC{{\it Nuov.~Cim.~}}
\def\SJNP{{\it Sov.~Jour.~Nucl.~Phys.~}}
\def\SPJETP{{\it Sov.~Phys.~J.E.T.P.~}}
\def\ZP{{\it Zeit.~Phys.~}}
\def\JP{{\it Jour.~Phys.~}}
\def\MPL{{\it Mod.~Phys.~Lett.~}}
\def\JHEP{{\it Jour.~High~Energy~Phys.~}}
\def\vol#1{{\bf #1}}
\def\vyp#1#2#3{\vol{#1} (#2) #3}

\thispagestyle{empty}
\vbox{\hfill CERN-TH/2000-332}
\vbox{\hfill Edinburgh 2000/23}
\begin{center}
\vskip 36pt
\centerline{\Large\bf Soft Resummation of Quark Anomalous Dimensions} 
\vskip 12pt
\centerline{\Large\bf and Coefficient Functions in \MS\ Factorization} 
\vskip 72pt
\centerline{\Large Simon Albino$^{\ast}$ and Richard D. Ball$^\dagger$ }
\vskip 18pt
\centerline{\it Department of Physics and Astronomy}
\centerline{\it University of Edinburgh, EH9 3JZ, Scotland$^{\ast,\dagger}$}
\vskip 18pt
\centerline{and}
\vskip 18pt
\centerline{\it Theory Division, CERN}
\centerline{\it CH-1211 Geneva 23, Switzerland$^\dagger$}
\vskip 120 pt
\centerline{\bf Abstract}
{\narrower\baselineskip 10pt
\medskip\noindent
The asymptotic behaviour at large $N$ of the \MS\ quark anomalous
dimensions is derived to all orders assuming only \MS\ factorization and 
standard results for the exponentiation of soft logarithms in the quark initiated 
bare cross sections for deep inelastic scattering and Drell-Yan. 
The result is then used to write the \MS\ quark coefficient functions 
in a form in which all terms of $O(\ln^{m} N)$ are resummed.}
\end{center}
\vfill
{November 2000\hfill}\\

\vfill\eject
\addtocounter{page}{-1}
Quark initiated bare cross sections, whether for deep inelastic or Drell-Yan processes,  
generally contain logarithmic singularities as $z\to 1$, where $z$ 
is the longitudinal momentum fraction of the participating partons. These logarithms result
from soft and collinear gluon emission, and in Mellin space result in 
logarithms of the form $\ln^{m} N$, which can be shown to exponentiate to all orders in 
perturbation theory \cite{A6,C1}.
Much progress has been made \cite{A6}-\cite{Y} in resumming these logarithms.
Once the general form of the exponent is determined, individual coefficients are fixed 
through matching to fixed order perturbation theory.

Before comparing to large $z$ data it is first necessary to resum the logarithms in the 
{\it factorized} quark initiated cross sections, the quark coefficient functions. 
The resummation is then expected to improve the convergence of the perturbation series
at large $N$. In order to factorize the cross section it is generally assumed (see for example 
ref. \cite{A43}) that in \MS\ factorization no resummation is necessary for the quark 
anomalous dimensions, i.e. that higher order contributions to the anomalous dimensions grow 
no faster than the leading order contribution at large $N$.
Considerable support for this assumption was given in ref. \cite{A30}, where the 
large $N$ behaviour of the \MS\ anomalous dimensions was determined to all orders
using eikonal techniques.

In this letter we will show that the all order exponentiation \cite{A6,C1} and \MS\ factorization
are by themselves sufficient to determine the all order large $N$ behaviour of the quark 
anomalous dimensions.  More specifically, we take the large $N$ resummed form of the 
quark initiated bare cross sections obtained in ref. \cite{C1}, then apply dimensional regularization
in order to factorize the infrared singularities in 
\MS\ scheme (ref. \cite{C1} uses an explicit infrared cutoff).
Matching the bare and factorized cross sections then allows us to simply read off the 
asymptotic behaviour of the quark anomalous dimensions: we find that the 
$\overline{MS}$ nonsinglet $qq$ anomalous dimension $\gamma^{NS}_{qq}$ behaves
like $O(\ln N)$ at large $N$, whereas $\gamma^{NS}_{q\bar{q}}$ behaves
like $O\left(1/N\right)$. Finally, we obtain a general form for the 
$\overline{MS}$ quark coefficient functions at large $N$ to all
orders. The results are then illustrated by explicit computation at LL and NLL.

{\bf 1.}~We consider DIS and Drell-Yan bare quark initiated cross sections, which
after factoring off all electroweak factors and decomposing into Lorentz invariants,
may be written as functions of two kinematic variables, $Q^2$ and $z$.
For DIS $Q^2$ is the virtuality of the 
incoming vector boson, while in DY it is the mass of the outgoing dileptons. Similarly,  
the longitudinal momentum fraction $z=x=Q^2/2p.q$ for DIS, while in DY $z = x_1x_2 = 
Q^2/s$. In both cases we will assume all quarks are 
massless, and use $a_s = \alpha_s/2\pi$ as a perturbative expansion parameter,
ignoring all contributions which are suppressed by powers of $Q^2$.

Now soft and collinear gluons radiated from the incoming quark lines can generate terms  
of the form $a_{s}^n\left[\frac{\ln^{m-1}(1-z)}{1-z}\right]_{+}$, $1\leq m \leq 2n$,
which diverge as $z\to1$, and which thus need to be resummed if the 
perturbative cross section is to be improved at large $z$. In Mellin space, 
these terms become

\begin{equation}
\int_{0}^{1}dzz^{N-1}\left[\frac{\ln^{m-1}(1-z)}{1-z}\right]_{+}=
\int_{0}^{1}dz\frac{z^{N-1}-1}{1-z}\ln^{m-1}(1-z)
=\sum_{i=0}^{m}b_{i}^{m}\ln^{i} N+O\left(\frac{1}{N}\right),
\label{Mellin}
\end{equation}

so the singularities as $z \rightarrow 1$ become logarithmic singularities
as $N \rightarrow \infty$. Note that in the remainder term we do not 
distinguish terms of $O\left(\frac{\ln^{m}N}{N}\right)$ from
terms of $O\left(\frac{1}{N}\right)$. In $z$ space there are also 
terms in the bare cross section from virtual graphs 
of the form $\delta(1-z)$, which in $N$ space are constant at large $N$.
However all the diagrams in which the initial quark (or at least 
one of the two initial quarks in the case of Drell-Yan) is not connected to
the electroweak boson via a single quark line, i.e. the diagrams
contributing to a bare pure singlet cross section, are either non 
singular or of the form $\ln^{m}(1-z)$ as $z \rightarrow 1$, and are thus of 
$O(1/N)$ after Mellin transformation. Here we will thus be concerned only
with the nonsinglet diagrams, since it is only these  which can have large 
$N$ singularities of the form in eqn.~(\ref{Mellin}). Furthermore, we will only be concerned with the 
singular parts of these diagrams (or more precisely those parts which at large $N$ are of 
$O(\ln^{m} N)$, $m=0,...,\infty$), which are independent of the type or polarization 
of the electroweak boson (and thus for example in DIS will be identical for all three nonsinglet 
structure functions $F_{1,2,3}$, and in DY for dimuon or $W$ production, after removing electroweak
factors). 

Consequently, here we need consider only two partonic cross sections 
$\sigma^{[a]}_{q}\left(z,{Q^{2}}/{\kappa^{2}},a_{s}(Q^{2})\right)$,
where the only index $a$ denotes the number of initial state quark lines: 
$a=1$ for DIS, and $a=2$ for DY. Both partonic cross sections are normalized 
such that $\sigma^{[a]}_{q}=1$ for $a_{s}=0$. Collinear singularities
are regulated by a generic infrared cutoff $\kappa$. 
In general \cite{A6,D1}, the perturbation series for ${\sigma}^{[a]}_{q}$
in $N$ space is then (given eqn.~(\ref{Mellin})) of the form

\begin{equation}
{\sigma}^{[a]}_{q}\left(N,\frac{Q^{2}}{\kappa^{2}},a_{s}(Q^{2})\right)
=1+\sum_{n=1}^{\infty}a_{s}^{n}(Q^{2})
\sum_{m=0}^{2n}c^{[a]}_{nm}\left({Q^{2}}/{\kappa^{2}}\right)
\ln^{m}N+O\left(\frac{1}{N}\right),
\label{F2LNlargeN}
\end{equation}

for some coefficients $c^{[a]}_{nm}\left({Q^{2}}/{\kappa^{2}}\right)$. 
Furthermore, the $O(\ln^{m} N)$ terms, for $m=0,...,\infty$, not only factorize:
in Mellin space they also exponentiate, that is the bare cross section may be 
written in the form 

\begin{eqnarray}
&&\ln{\sigma}^{[a]}_{q}(N,Q^{2}/\kappa^{2},a_{s}(Q^{2}))
=\phi_{-1}^{[a]}\big(a_{s}(Q^{2})\ln N,Q^{2}/\kappa^{2}\big)\ln N \cr
&&\qquad\qquad\qquad\qquad+\sum_{n=0}^{\infty}(a_{s}(Q^{2}))^{n}
\phi_{n}^{[a]}\big(a_{s}(Q^{2})\ln N,Q^{2}/\kappa^{2}\big)
+O\left(\frac{1}{N}\right).
\label{genexpF2LN}
\end{eqnarray}

This exponentiation provides a convenient framework for organising the expansion:
the functions $\phi^{[a]}_{-1}$ contains all the 
leading logarithms (LL), $\phi^{[a]}_{0}$
the next-to-leading logarithms (NLL), since there is an extra power of $a_s$, and so on.
It is easy to see that in eqn.~(\ref{F2LNlargeN}) the LL are then those terms with $n+1 \leq m \leq 2n$,
NLL, those with $m=n$, etc. Terms of $O\left(\frac{1}{N}\right)$ in ${\sigma}^{[a]}_{q}$
will be systematically discarded in what follows, since they are not included in its definition.

To complete the resummation, we must determine the functions $\phi^{[a]}_i$. This 
may be achieved either by 
eikonal arguments \cite{A6,C1,D1}, or by applying the renormalization group directly to the 
factorization of the large $N$ singularities \cite{A17}. The result is rather simple: 
the resummation may be performed in closed form essentially through a change in the argument 
of the running coupling from $a_s(Q^2)$ to $a_s((1-z)^a Q^2)$. Specifically
it is found that (see for example ref.~\cite{D1}) 

\begin{eqnarray}
\ln {\sigma}^{[a]}_{q}
=\int_{0}^{1}dz\frac{z^{N-1}-1}{1-z}
\bigg( a\int_{\kappa^{2}}^{(1-z)Q^{2}}\frac{dq^{2}}{q^{2}}A(a_{s}(q^{2}))
&+&B^{[a]}(a_{s}((1-z)Q^{2}))\bigg)\cr
&&+K^{[a]}(a_{s}(Q^{2})),
\label{sigLLNLL}
\end{eqnarray}

where the $\overline{MS}$ renormalization scheme has been used, and soft singularities are 
regulated by the infrared cutoff $\kappa$. The functions $A(a_s)$, $B^{[a]}(a_s)$, $R^{[a]}(a_s)$
each have perturbative expansions beginning at $O(a_s)$. For DY processes the second 
term is absent, so $B^{[2]}(a_s)=0$: in ref. \cite{D1} $B^{[1]}(a_s)$ is simply denoted by $B(a_s)$. 
The remainder terms $K^{[a]}(a_s)$ contain contributions not necessarily related to 
soft or collinear gluons. To resum LL and NLL logarithms only the first two 
coefficients in the expansion of $A$, and the first coefficient in the expansion of $B^{[1]}$
are necessary, and these may be read off by making a direct comparison to the usual fixed order 
LO and NLO cross sections expanded at large $N$.

In order to express the resummation eqn.~(\ref{sigLLNLL}) in a form 
more amenable to subsequent discussion, we first rewrite them in $d=4-2\epsilon$ 
dimensions. We may remove the collinear regulator by taking $\kappa \rightarrow 0$, 
since bare cross sections are non singular in a non integer number of dimensions.
This gives

\begin{eqnarray}
\ln {\sigma}^{[a]}_{q}=\int_{0}^{1}dz \frac{z^{N-1}-1}{1-z}
\bigg(a\int_{0}^{(1-z)Q^{2}}\frac{dq^{2}}{q^{2}}
A(a_{s}(q^{2},\epsilon),\epsilon)&+&
B^{[a]}(a_{s}((1-z)Q^{2},\epsilon),\epsilon)\bigg)\cr
&&+K^{[a]}(a_{s}(Q^{2}),\epsilon)+O(\epsilon),
\label{R1indimreg}
\end{eqnarray}

for the DIS and DY cross sections respectively. Here $a_{s}(\mu^{2},\epsilon)$ is the \MS\ 
renormalised coupling in $4-2\epsilon$ dimensions: it satisfies the 
renormalization group equation 

\begin{equation}
\frac{\partial a_{s}(\mu^{2},\epsilon)}{\partial \ln \mu^{2}}
=-\epsilon a_{s}(\mu^{2},\epsilon)-\sum_{n=0}^{\infty}\beta_{n}
a_{s}^{n+2}(\mu^{2},\epsilon),
\label{betaeqindimreg}
\end{equation}

where the $\beta$-function coefficients $\beta_{n}$ are independent of $\epsilon$.
Clearly $a_{s}(\mu^{2},0)=a_{s}(\mu^{2})$, while $a_{s}(0,\epsilon)=0$ (by 
analytic continuation from $\epsilon<0$). The functions                
$A(a_{s},\epsilon)$, $B^{[a]}(a_{s},\epsilon)$ and $K^{[a]}(a_{s},\epsilon)$ all have 
perturbative expansions in powers of $a_s$, but now the coefficients in these 
expansions will depend on $\epsilon$. The terms of $O(\epsilon)$ in eqn.~(\ref{R1indimreg}) are 
inconsequential and can be dropped.

Now using eqn.~(\ref{betaeqindimreg}) for the running of  $a_{s}(q^{2},\epsilon)$, we 
can do the $q^{2}$ integrals term in an expansion in powers of $a_s$:

\begin{equation}
\int_{0}^{Q^2}{d\ln q^{2}}(a_{s}(q^{2},\epsilon))^{n}=
\int_{0}^{a_{s}(Q^{2},\epsilon)}d a_{s} \frac{a_{s}^{n}}{-\epsilon a_{s}-\sum_{j=0}^{\infty}
\beta_{j} a_{s}^{j+2}}.
\label{intofasnintermsofasq2}
\end{equation}

Then from eqn.~(\ref{R1indimreg}), we find that the quark initiated bare cross sections 
at large $N$ are of the form

\begin{eqnarray}
\ln \sigma^{[a]}_{q}(N,a_{s}(Q^{2},\epsilon),\epsilon)
&=&\int_{0}^{1}dz \frac{z^{N-1}-1}{1-z}\left[
\sum_{i=1}^{\infty}
f^{[a]}_i(\epsilon)(a_{s}((1-z)^{a}Q^{2},\epsilon))^{i}\right] \cr
&&\qquad\qquad\qquad+\sum_{i=1}^{\infty}g^{[a]}_i(\epsilon)(a_{s}(Q^{2},\epsilon))^{i}
+O(\epsilon).
\label{genresoffhattoa2}
\end{eqnarray}

As advertised above, we can now see explicitly the essential feature of the 
large $N$ resummation: the change in the 
argument of the running coupling from $Q^2$ to $(1-z)^aQ^2$.


{\bf 2.}~In this section 
we will factorize eqn.~(\ref{genresoffhattoa2}) in the \MS\ scheme.
The factorization procedure involves separating out the collinear singularities 
from the bare cross sections $\sigma^{[a]}_{q}$ into a universal singular factor
$\Gamma_q$:

\begin{equation}
\ln \sigma_{q}^{[a]}=\ln {C}_{q}^{[a]} + a\ln \Gamma_{q}.
\label{lnfactofnonsingforDISplusDY}
\end{equation}

The $C_{q}^{[a]}$ are process dependent coefficient functions, non singular as
$\epsilon \rightarrow 0$. Just as the complete bare partonic cross sections 
for nonsinglet, singlet and valence processes are all proportional to 
$\sigma_{q}^{[a]}$ up to terms of $O(1/N)$, so the complete nonsinglet, 
singlet and valence coefficient functions will all be proportional to
${C}_{q}^{[a]}$ up to terms of $O(1/N)$. Of course the same holds true for the 
singular factor $\Gamma_q$, and in particular for the anomalous dimensions
from which it is constructed: in \MS\ $\Gamma_q$ satisfies the renormalization 
group equation

\begin{equation}
\frac{\partial\ln \Gamma_q(N,a_{s}(\mu^2,\epsilon),\epsilon)}{\partial\ln\mu^{2}} =
\gamma_q(N,a_{s}(\mu^2,\epsilon)),
\label{diffofgamintoP}
\end{equation}

where $\gamma_q(N,a_{s})=\sum_{n=0}^{\infty}\gamma_q^{n}(N)a_{s}^{n}$ is the 
anomalous dimension, and in the \MS\ scheme the coefficients $\gamma_q^{n}(N)$ are independent
of $\epsilon$. This defines the \MS\ factorization scheme. This equation has the usual solution

\begin{equation}
\Gamma_q(N,a_{s}(\mu^{2},\epsilon),\epsilon)
= \exp\left[\int_{0}^{\mu^{2}}{d\ln q^{2}}\gamma_q(N,a_{s}(q^{2},\epsilon))
\right],
\label{Gamindimreg}
\end{equation}

which, when combined with eqn.~(\ref{betaeqindimreg}) and expanded in powers of $a_s$
generates all the collinear singularities of the bare cross section, 
in the form of inverse powers of $\epsilon$.

We first use eqn.~(\ref{lnfactofnonsingforDISplusDY}) to determine 
the degree of divergence of the coefficients
$f^{[a]}_i(\epsilon)$ and $g^{[a]}_i(\epsilon)$ 
as $\epsilon \rightarrow 0$. Differentiating eqn.~(\ref{lnfactofnonsingforDISplusDY}) with
respect to $\ln Q^{2}$, and using eqn.~(\ref{diffofgamintoP}), we find

\begin{equation}
\frac{\partial \ln {\sigma}^{[a]}_{q}}{\partial \ln Q^{2}}
=\frac{\partial\ln C_{q}^{[a]}}{\partial \ln Q^{2}}
+a\gamma_q(N,a_{s}(Q^{2},\epsilon)).
\end{equation}

Since both terms on the right hand side are nonsingular as $\epsilon \rightarrow 0$, 
it follows that ${\partial\ln \sigma_{q}^{[a]}}/{\partial\ln Q^{2}}$ is non singular. 
Differentiation of eqn.~(\ref{genresoffhattoa2}) then leads to the conclusion that

\begin{equation}
f^{[a]}_i(\epsilon)=\sum_{t=0}^{i}f^{[a]}_{i,t}\epsilon^{-t}+O(\epsilon),
\qquad
g^{[a]}_{i}(\epsilon)=\sum_{t=0}^{i}g^{[a]}_{i,t}\epsilon^{-t}+O(\epsilon),
\end{equation}

as well as various relations among the $f^{[a]}_{i,t}$, and $g^{[a]}_{i,t}$.

Using eqn.~(\ref{Gamindimreg}), we can rewrite $\ln \Gamma_{q}$ in the form 

\begin{equation}
\ln \Gamma_{q}(N,a_{s}(Q^{2},\epsilon),\epsilon) 
=\int_{0}^{1}dz z^{N-1}\int_{0}^{Q^2}\frac{dq^2}{q^2}
P_{q}(z,a_{s}(q^2,\epsilon)).
\label{gamMSsuit}
\end{equation}

where $P_{q}(z,a_{s})=\sum_{n=1}^\infty P_q^n(z)a_s^i$ is the quark 
splitting function (so the $P_q^n(z)$  Mellin transform to the anomalous 
dimensions $\gamma_q^n$). Since the $\gamma_q^{n}(N)$ are independent of $\epsilon$, so too
are the $P_{q}^i(z)$.
Substituting eqns.~(\ref{gamMSsuit},\ref{genresoffhattoa2}) into 
eqn.~(\ref{lnfactofnonsingforDISplusDY}) we then find that

\begin{eqnarray}
\ln C_{q}^{[a]}&=&\int_{0}^{1}dz (z^{N-1}-1)\Bigg[\frac{1}{1-z}\sum_{i=1}^{\infty}
\sum_{t=1}^{i}f^{[a]}_{i,t}\epsilon^{-t}(a_{s}((1-z)^{a}Q^{2},\epsilon))^{i}\cr
&&\qquad\qquad\qquad\qquad -a\int_{0}^{(1-z)^{a}Q^2}\frac{dq^2}{q^2}
P_q(z,a_{s}(q^2,\epsilon))
\Bigg]\cr
&& \qquad  +\left[\sum_{i=1}^{\infty}\sum_{t=1}^{i}g^{[a]}_{i,t}\epsilon^{-t} (a_{s}(Q^{2},\epsilon))^{i}
-a\int_{0}^{1}dz\int_{0}^{Q^2}\frac{dq^2}{q^2}P_q(z,a_{s}(q^{2},\epsilon))\right]
\cr
&& \qquad
+\Bigg[\int_{0}^{1}dz (z^{N-1}-1)\frac{1}{1-z}\sum_{i=1}^{\infty}
f^{[a]}_{i,0}(a_{s}((1-z)^{a}Q^{2},\epsilon))^{i}
+\sum_{i=1}^{\infty}g^{[a]}_{i,0}(a_{s}(Q^{2},\epsilon))^{i}
\cr
&& \qquad
-a\int_{0}^{1}dz (z^{N-1}-1)\int_{(1-z)^{a}Q^2}^{Q^2}\frac{dq^2}{q^2}
P_q(z,a_{s}(q^2,\epsilon))\Bigg]+O\left(\frac{1}{N}\right)+O(\epsilon).
\label{ftildeatlargeNgen}
\end{eqnarray}

The motivation for organisation of the terms in this equation is that, as we will now show,
each of the three pairs of square brackets is finite as $\epsilon \rightarrow 0$. 

We first work with the first pair of
square brackets. Using eqn.~(\ref{betaeqindimreg}), we can perform the integration of the 
$a_{s}^{n}(q^2)$ in the second term as

\begin{eqnarray}
\int_{0}^{(1-z)^{a}Q^2}\frac{dq^2}{q^2}(a_{s}(q^2,\epsilon))^{n}&=&
\int_{0}^{a_{s}((1-z)^{a} Q^{2},\epsilon)}d a_{s} \frac{a_{s}^{n}}{-\epsilon a_{s}-\sum_{j=0}^{\infty}
\beta_{j} a_{s}^{j+2}}
\cr
&=&\sum_{i=n}^{\infty}p_{i,n}(\epsilon)(a_{s}((1-z)^{a} Q^{2},\epsilon))^{i}
\label{intofasintheps}.
\end{eqnarray}

It follows that

\begin{eqnarray}
\int_{0}^{(1-z)^{a}Q^2} \frac{dq^2}{q^2} 
P_q(z,a_{s}(q^{2},\epsilon))
&=&\sum_{n=1}^{\infty}\int_{0}^{(1-z)^{a}Q^2}\frac{dq^2}{q^2} 
(a_{s}(q^2,\epsilon))^{n} P_{q}^{n}(z)\cr
&=&\sum_{n=1}^{\infty}\sum_{i=n}^{\infty}P_{q}^{n}(z)
p_{i,n}(\epsilon)(a_{s}((1-z)^{a} Q^{2},\epsilon))^{i}\cr
&=&\sum_{i=1}^{\infty}\sum_{n=1}^{i}P_{q}^{n}(z)p_{i,n}(\epsilon)
(a_{s}((1-z)^{a} Q^{2},\epsilon))^{i}.
\label{intotoomzaofPinasomzaQ2}
\end{eqnarray}

We now expand eqn.~(\ref{intotoomzaofPinasomzaQ2}) in $\epsilon$. 
In eqn.~(\ref{intofasintheps}), the $p_{i,n}(\epsilon)$ may be expanded
as

\begin{equation}
p_{i,n}(\epsilon)=\sum_{s=1}^{i-n+1}p^{s}_{i,n}\epsilon^{-s}.
\end{equation}

Note that $p^{s}_{i,n}=0$ if $s=1$ and $i \neq n$, but we do not explicitly show this
in order to simplify the notation. Substitution in eqn.~(\ref{intotoomzaofPinasomzaQ2})
then gives finally

\begin{eqnarray}
\int_{0}^{(1-z)^{a}Q^2} \frac{dq^2}{q^2} 
P_q(z,a_{s}(q^{2},\epsilon))
&=&\sum_{i=1}^{\infty}\sum_{n=1}^{i}\sum_{s=1}^{i-n+1}P_q^{n}(z)p^{s}_{i,n}\epsilon^{-s}
(a_{s}((1-z)^{a} Q^{2},\epsilon))^{i}\cr
&=&\sum_{i=1}^{\infty}\sum_{t=1}^{i}\sum_{n=1}^{i-t+1}P_q^{n}(z)p^{t}_{i,n}\epsilon^{-t}
(a_{s}((1-z)^{a} Q^{2},\epsilon))^{i}.
\label{expanofPintinepandas}
\end{eqnarray}

The second term in the second pair of square brackets in 
eqn.~(\ref{ftildeatlargeNgen}) may be expanded similarly, using  
eqn.~(\ref{expanofPintinepandas}) with $a=0$:

\begin{equation}
\int_{0}^{Q^2} \frac{dq^2}{q^2} 
P_q(z,a_{s}(q^{2},\epsilon))
=\sum_{i=1}^{\infty}\sum_{t=1}^{i}\sum_{n=1}^{i-t+1}P_q^{n}(z)
p^{t}_{i,n}\epsilon^{-t}
(a_{s}(Q^{2},\epsilon))^{i}.
\label{expanofPintinepandasazero}
\end{equation}

Finally, the last term in the third pair of brackets in 
eqn.~(\ref{ftildeatlargeNgen}) is given by the difference of 
eqns.~(\ref{expanofPintinepandas}) and (\ref{expanofPintinepandasazero}), and is thus 
finite as $\epsilon \rightarrow 0$. Indeed, we can write it as

\begin{equation}
\int^{Q^2}_{(1-z)^{a}Q^2}\frac{dq^2}{q^2}
P_q(z,a_{s}(q^{2},\epsilon))
=\int^{Q^{2}}_{(1-z)^{a} Q^{2}}
\frac{d q^2}{q^2}\sum_{n=1}^{\infty}(a_{s}(q^2))^{n}P_q^{n}(z)+O(\epsilon).
\label{thirdbit}
\end{equation}

Putting all this together, substituting eqns.~(\ref{expanofPintinepandas},
\ref{expanofPintinepandasazero},\ref{thirdbit}) into eqn.~(\ref{ftildeatlargeNgen}) 
we find

\begin{eqnarray}
\ln C_{q}^{[a]}&=&\int_{0}^{1}dz (z^{N-1}-1)\Bigg[\frac{1}{1-z}\sum_{i=1}^{\infty}
\sum_{t=1}^{i}f^{[a]}_{i,t}\epsilon^{-t}(a_{s}((1-z)^{a}Q^{2},\epsilon))^{i}
\cr
&& \qquad\qquad\qquad\qquad
-a\sum_{i=1}^{\infty}\sum_{t=1}^{i}\sum_{n=1}^{i-t+1}P_q^{n}(z)p^{t}_{i,n}\epsilon^{-t}
(a_{s}((1-z)^{a} Q^{2},\epsilon))^{i}\Bigg]
\cr
&& \qquad
+\left[\sum_{i=1}^{\infty}\sum_{t=1}^{i}g^{[a]}_{i,t}\epsilon^{-t}
(a_{s}(Q^{2},\epsilon))^{i}
-a\int_{0}^{1}dz\sum_{i=1}^{\infty}\sum_{t=1}^{i}\sum_{n=1}^{i-t+1}
P_q^{n}(z)p^{t}_{i,n}\epsilon^{-t}
(a_{s}(Q^{2},\epsilon))^{i}\right]
\cr
&& \qquad
+\Bigg[\int_{0}^{1}dz (z^{N-1}-1)\frac{1}{1-z}\sum_{i=1}^{\infty}
f^{[a]}_{i,0}(a_{s}((1-z)^{a}Q^{2},\epsilon))^{i}
+\sum_{i=1}^{\infty}g^{[a]}_{i,0}(a_{s}(Q^{2},\epsilon))^{i}
\cr
&& \qquad
-a\int_{0}^{1}dz (z^{N-1}-1)\int^{Q^{2}}_{(1-z)^{a} Q^{2}}
\frac{d q^2}{q^2}\sum_{n=1}^{\infty}(a_{s}(q^2))^{n}P_q^{n}(z)\Bigg]+O(\epsilon).
\label{ftildeatlargeNgen2}
\end{eqnarray}

Now, consider only the coefficients of $\epsilon^{-t}$ for $t=1,...,\infty$ 
in eqn.~(\ref{ftildeatlargeNgen2}),
which must vanish so that $C_{q}^{[a]}$ is finite in the limit $\epsilon \rightarrow 0$.
In the first pair of square brackets, these coefficients are all functions of $N$, in the
second pair of square brackets they are all independent of $N$, while 
all terms in the third pair of square brackets are nonsingular. Thus
each of the three pairs of square brackets is separately nonsingular as $\epsilon \rightarrow 0$.
Moreover the cancellation of singularities in the first pair of square brackets implies that,
for $i \geq t \geq 1$ and $z \neq 1$,

\begin{equation}
\frac{1}{1-z}f^{[a]}_{i,t}=a\sum_{n=1}^{i-t+1}P_q^{n}(z)p^{t}_{i,n},
\label{relofPandW}
\end{equation}

We can treat $p^{t}_{i,n}$ in eqn.~(\ref{relofPandW}) as a set of matrices with indices 
$i$ and $n$, but vanishing for $n>i-t+1$; each matrix is then triangular and thus invertible. 
Then multiplication of both sides of eqn.~(\ref{relofPandW}) by the inverse of this matrix gives an expression
with just $P_q^{n}$ on the right hand side, with the left hand side proportional
to $\left(1/(1-z)\right)$. Thus we find that for $z\neq 1$

\begin{equation}
P_q(z,a_{s})=Q(a_{s})\frac{1}{1-z}.
\end{equation}

where $Q(a_s)=\sum_{i=1}^\infty a_s^iQ_i$, and the $Q_i$ may be determined by substitution into 
eqn.~(\ref{relofPandW}). Note that for consistency we also require that $f^{[a]}_{i,t}$ is 
proportional to $a$ for $t>0$: this is the essence of universal factorization.
Similarly the cancellation of singularities in the second bracket 
implies that

\begin{equation}
g^{[a]}_{i,t}=a\sum_{n=1}^{i-t+1}\int_0^1dz P_q^{n}(z)p^{t}_{i,n},
\label{relofPandWsec}
\end{equation}

and thus that $\int_0^1dzP_q^{n}(z)$ is finite. It follows that for all $z$, 
we may write

\begin{equation}
P_q(z,a_{s})=Q(a_{s})\left[\frac{1}{1-z}\right]_{+}
+R(a_{s})\delta(1-z),
\label{mainresultP}
\end{equation}

where again $R(a_s)=\sum_{i=1}^\infty a_s^i R_i$, and the $R_i$ may be determined by
substituting eqn.~(\ref{mainresultP}) into eqn.~(\ref{relofPandWsec}) and inverting.

Thus we find that, in Mellin space, as $N \rightarrow \infty$, the anomalous dimension

\begin{equation}
\gamma_q(N,a_s)=-Q(a_s)(\ln N+\gamma_E) + R(a_s)+O(1/N),
\label{asympofPplus}
\end{equation}

where $\gamma_E$ is Euler's constant. Remembering that all the nonsinglet, 
singlet and valence anomalous dimensions are equal to 
$\gamma_q$ up to terms which vanish as $1/N$ in the large $N$ limit,
then since in the usual decomposition

\begin{equation}
\gamma_{q_iq_j} = \delta_{ij}\gamma^{NS}_{qq}+\gamma^{PS}_{qq},
\qquad \gamma_{q_i\bar{q}_j} = \delta_{ij}\gamma^{NS}_{q\bar{q}}+\gamma^{PS}_{q\bar{q}},
\qquad \gamma^{NS}_{\pm}=\gamma^{NS}_{qq}\pm\gamma^{NS}_{q\bar{q}},
\end{equation}

we must have

\begin{eqnarray}
&&\gamma^{NS}_{qq}(N) = \gamma_q(N)+O(1/N),\qquad\qquad\gamma^{NS}_{q\bar{q}}(N)=O(1/N), \cr
&&\gamma^{PS}_{qq}(N)=O(1/N),\qquad\qquad\qquad\qquad \gamma^{PS}_{q\bar{q}}(N)=O(1/N),
\end{eqnarray}

consistent with our remarks earlier that diagrams in which the quark evolves into a gluon are 
suppressed by $1/N$. 

It remains to take the limit $\epsilon \rightarrow 0$ in eqn.~(\ref{ftildeatlargeNgen2}) to 
give an explicit expression for the resummed large $N$ \MS\ coefficient function: we find

\begin{eqnarray}
\ln C_{q}^{[a]}
&=&\int_{0}^{1}dz \frac{z^{N-1}-1}{1-z}\Big[a\int_{Q^{2}}^{(1-z)^{a} Q^{2}}
\frac{d q^2}{q^2} \sum_{n=1}^{\infty}Q_{n}(a_{s}(q^2))^{n}+
\sum_{i=1}^{\infty}
f^{[a]}_{i,0}(a_{s}((1-z)^{a}Q^{2}))^{i}
\Big]\cr
&&\qquad\qquad\qquad\qquad +\sum_{i=1}^{\infty}g^{[a]}_{i,0}(a_{s}(Q^{2}))^{i}
+O\left(\frac{1}{N}\right).
\label{ftildeatlargeNgen3}
\end{eqnarray}

We immediately recognise this as eqn.~(\ref{sigLLNLL}), with

\begin{equation}
A(a_s) =\sum_{n=1}^{\infty}Q_{i}a_{s}^{i},\qquad
B^{[a]}(a_s) = \sum_{i=1}^{\infty}f^{[a]}_{i,0}a_s^{i},\qquad
K^{[a]}(a_s) = \sum_{i=1}^{\infty}g^{[a]}_{i,0}a_{s}^{i},
\end{equation}

and with the infrared regulator set equal to 
the factorization scale, i.e. $\kappa^2=Q^2$.

To calculate the coefficients $f^{[a]}_{i,0}$ and the 
$g^{[a]}_{i,0}$, we simply expand eqn.~(\ref{ftildeatlargeNgen3}) 
in $a_{s}(Q^{2})$, i.e. undo the resummation, and compare coefficients 
of $a_{s}^{i}(Q^{2})$ with those in the fixed order coefficient functions. The universal 
coefficients $Q_{i}$ of the $O(\ln N)$ terms in the anomalous dimension serve as a 
consistency check. This 
procedure will be illustrated explicitly for all LL and NLL terms in the next section.


{\bf 3.}~We now show explicitly that the factorized coefficient functions eqn.~(\ref{ftildeatlargeNgen3}) 
correctly resum all LL and NLL terms, by comparing then to the fixed order NLO and NNLO 
coefficient functions and thereby deduce the leading behaviour at large $N$ of the LO and 
NLO quark anomalous dimensions. 

To NLO, eqn.~(\ref{ftildeatlargeNgen3}) reads

\begin{eqnarray}
&\ln& C_{q}^{[a]}(N,a_{s}(Q^{2}))=\int_{0}^{1}dz \frac{z^{N-1}-1}{1-z}
\bigg[-\frac{aQ_1}{\beta_{0}}\ln \left(\frac{a_{s}((1-z)^{a}Q^{2})}{a_{s}(Q^{2})}\right)
\cr
& \quad &
+a\left(\frac{Q_1\beta_{1}}{\beta_{0}^{2}}-\frac{Q_2}{\beta_{0}}\right)
(a_{s}((1-z)^{a}Q^{2})-a_{s}(Q^{2}))
+f^{[a]}_{1,0}a_{s}((1-z)^{a}Q^{2})\bigg]\cr
&\quad&\qquad\qquad\qquad\qquad\qquad
+O(a_{s}(a_{s}\ln N)^{m}).
\label{factcoefffuncresumgenNLO}
\end{eqnarray}

Performing the $z$ integral, we find

\begin{eqnarray}
\ln C_{q}^{[a]}(N,a_{s})
&=&\frac{Q_1}{\beta_{0}}\frac{1}{a_{s}\beta_{0}}
\left[(1-a\lambda_s)\ln (1-a\lambda_s)+a\lambda_s\right]
\cr
&& \qquad
+\left(\frac{f^{[a]}_{1,0}}{a\beta_{0}}
-\frac{aQ_1\gamma_{E}}{\beta_{0}}
+\frac{Q_1\beta_{1}}{\beta_{0}^{3}}-\frac{Q_2}{\beta_{0}^{2}}\right)
\ln (1-a\lambda_s)
\cr
&& \qquad
+\frac{Q_1\beta_{1}}{2 \beta_{0}^{3}}\ln^{2} (1-a\lambda_s)
-\left(\frac{Q_2}{\beta_{0}^{2}}
-\frac{Q_1\beta_{1}}{\beta_{0}^{3}}\right)a\lambda_s 
+O(a_{s}(a_{s}\ln N)^{m}),
\label{factcoefffuncresumgenNLO3}
\end{eqnarray}

where $\lambda_s\equiv a_{s}\beta_{0}\ln N$.
This result should be compared to the general expression eqn.~(\ref{genexpF2LN}): 
the first line gives the LL terms $\phi_{-1}$, while the rest gives the NLL $\phi_{0}$. 

To determine the large $N$ behaviour of the LO and NLO anomalous dimensions,
i.e. the coefficients $Q_1$ and $Q_2$, and also the nonuniversal coefficients $f^{[a]}_{1,0}$, 
it is sufficient to expand eqn.~(\ref{factcoefffuncresumgenNLO3})
in $a_s(Q^{2})$ to NNLO and compare to NL0 and NNLO coefficient functions at large $N$. 
The result, after exponentiating, is

\begin{eqnarray}
C_{q}^{[a]}&(&N,a_{s})
=1+a_{s}\left(\frac{1}{2}a^{2}Q_1\ln^{2} N+
\left(a^{2}Q_1\gamma_{E}-f^{[a]}_{1,0}\right)
\ln N\right)
\cr
&&
+a_{s}^{2}\Big(
\frac{1}{8}a^4Q_1^2\ln^{4} N
+\left[\frac{1}{6}a^{3}Q_1\beta_{0}+
+a^2Q_1\left(a^{2}Q_1\gamma_{E}
-f^{[a]}_{1,0}\right)\right]\ln^{3} N
\cr
&&
+\bigg[\frac{1}{2}a^{3}\beta_{0}Q_1\gamma_{E}-\frac{1}{2}a\beta_{0}
f^{[a]}_{1,0}+\frac{1}{2}a^{2}Q_2
+\frac{1}{2}\left(a^{2}Q_1\gamma_{E}-f^{[a]}_{1,0}\right)^2\bigg]\ln^{2} N \Big)+O(a_{s}^{3}).
\label{factcoefffuncresumgenNLOunres}
\end{eqnarray}

Now, for large $N$, to NLO~\cite{BBDM} and NNLO~\cite{A8} the DIS quark coefficient function 

\begin{eqnarray}
C_{q}^{[1]}&(&N,a_{s})=1+a_s\Big[C_{F}\ln^{2} N+C_{F}\left(2\gamma_{E}+\frac{3}{2}\right)\ln N
+\gamma_{E}^{2}
+\frac{3}{2}\gamma_{E}-\zeta(2)+O\left(\frac{1}{N}\right)\Big]
\cr
&&
+a_s^2\Big[\left(2C_{F}^2\gamma_{E}-\frac{2}{9}C_{F}T_{R}n_{f}+\frac{3}{2}C_{F}^2
+\frac{11}{18}C_{F}C_{A}\right)\ln^{3}N
\cr
&&
+\Big(\frac{11}{6}C_{F}C_{A}\gamma_{E}+\frac{9}{2}C_{F}^2\gamma_{E}-C_{F}C_{A}\zeta(2)
-\frac{27}{8}C_{F}^2-C_{F}^2\zeta(2)+\frac{367}{72}C_{F}C_{A}
\cr
&&
+3C_{F}^2\gamma_{E}^2
-\frac{2}{3}C_{F}T_{R}n_{f}\gamma_{E}-\frac{29}{18}C_{F}T_{R}n_{f}\Big)\ln^{2}N
+O(\ln N)\Big].
\label{NNLOsigqDISlargeN}
\end{eqnarray}

Comparing this with eqn.~(\ref{factcoefffuncresumgenNLOunres}) with $a=1$, we find

\begin{equation}
Q_1=2C_{F},\qquad
Q_2=C_{F} C_{A} \left(\frac{67}{9}-2 \zeta(2)\right)-\frac{20}{9} C_{F} T_{R} n_{f},
\label{tildePV1qqval}
\end{equation}

consistent with the large $N$ behaviour of the LO~\cite{GW} and NLO~\cite{CFP} anomalous 
dimensions. Furthermore the coefficient

\begin{equation}
f^{[1]}_{1,0}=-\frac{3}{2}C_{F},
\label{valueoff110}
\end{equation}

as in \cite{C1,L}. For Drell-Yan at large $N$, to NLO \cite{AEM} and NNLO \cite{A10}

\begin{eqnarray}
C_{q}^{[2]}&(&N,a_{s})
=1 + a_s\Big[4 C_{F} \ln^{2}N+8 C_{F} \gamma_{E} \ln N 
+4 C_{F} \gamma_{E}^{2}
+8 C_{F} \zeta(2)-8 C_{F}+O(1)\Big]
\cr
&&
 + a_s^2\Big[8 C_{F}^2 \ln^{4}N+\left(\frac{44}{9} C_{F} C_{A}-\frac{16}{9} C_{F} T_{R} n_{f}
+32 C_{F}^2 \gamma_{E}\right) \ln^{3}N
\cr
&&
+\Big(-\frac{40}{9} C_{F} T_{R} n_{f}-\frac{16}{3} C_{F} T_{R} n_{f} \gamma_{E}
+\frac{44}{3} C_{F} C_{A} \gamma_{E}
+\frac{134}{9} C_{F} C_{A}-4 C_{F} C_{A} \zeta(2)\cr
&&
+32 C_{F}^2 \zeta(2)+48 C_{F}^2 \gamma_{E}^2-32 C_{F}^2\Big) \ln^{2}N
+O(\ln N)\Big],
\label{NNLOsigqDYlargeN}
\end{eqnarray}

which is again consistent with  eqn.~(\ref{factcoefffuncresumgenNLOunres}), this 
time with $a=2$, provided eqns.~(\ref{tildePV1qqval}) for the 
large $N$ anomalous dimensions are satisfied, and $f^{[2]}_{1,0}=0$. With this result and eqn.
(\ref{valueoff110}), eqn. (\ref{factcoefffuncresumgenNLO3}) agrees with the results
of \cite{A43} at NLL. Moreover, our general proof of eqn.(\ref{mainresultP})
now places the NNLL results of \cite{A43} on a firmer footing.


{\bf 4.}~We have used the dimensionally regularized form of the large $N$ resummed bare 
quark initiated cross sections
for DIS and Drell-Yan, eqn.~(\ref{genresoffhattoa2}), together with the fact that
the bare quark initiated cross sections are independent of the species of the electroweak boson,
and the definition of $\overline{MS}$ factorization, to obtain to all orders the $O(\ln N)$ behaviour
of the nonsinglet \MS\ anomalous dimensions $P^{NS}_{qq}$ and 
the $O\left(\frac{1}{N}\right)$ behaviour of $P^{NS}_{q\bar{q}}$ and the pure singlet.
This has interesting implications for the large $x$ evolution of \MS\ quark parton distribution 
functions: at large $N$ the evolution factor
\begin{equation}
\Gamma(Q^2/\mu^2)
=N^{-\int_{\mu^2}^{Q^2}\frac{d q^2}{q^2}Q(a_s(q^2))}(1+O(1/N)).
\label{GamlargeN}
\end{equation}
so that if $q(x,\mu^2)\sim (1-x)^{b(\mu^2)}$ as $x\to 1$, this behaviour persists
at higher scales with 
\begin{equation}
b(Q^2) = b(\mu^2) + \int_{\mu^2}^{Q^2}\frac{dq^2}{q^2}Q(a_s(q^2))
\end{equation}
order by order in perturbation theory. 

We then used the result (\ref{mainresultP}) to construct a general large 
$N$ resummed expression for the $\overline{MS}$ quark 
coefficient functions for DIS and DY, eqn.~(\ref{ftildeatlargeNgen3}), and in particular 
showed that all the large $N$ singularities in the DY case can be deduced from the $O(\ln N)$
terms in the quark anomalous dimension. We verified these results explicitly at LL and NLL.
For the future, the large $N$ behaviour of NNLO anomalous dimensions 
and NNNLO DIS and DY coefficient functions will provide a useful consistency check on new 
calculations.

\bigskip

{\bf Acknowledgements:} SDA would like to thank G. Sterman for very useful discussions on this topic, and 
W. van Neerven and S. Catani for correspondence.

\end{document}